\documentclass[aps,showpacs,twocolumn,showkeys,amsmath,amssymb,floatfix,nofootinbib,preprintnumbers]
{revtex4}
\pdfoutput=1
\usepackage{bm}
\usepackage{epsfig}

\begin{document}

\title{Do the $P_c^+$ Pentaquarks Have Strange Siblings?}

\author{Richard F. Lebed}
\email{richard.lebed@asu.edu}
\affiliation{Department of Physics, Arizona State University, Tempe,
Arizona 85287-1504, USA}

\date{October, 2015}

\begin{abstract}
  The recent LHCb discovery of states $P_c^+(4380)$, $P_c^+(4450)$,
  believed to be $c\bar c uud$ pentaquark resonances, begs the
  question of whether equivalent states with $c\bar c \to s\bar s$
  exist, and how they might be produced.  The precise analogue to the
  $P_c^+$ discovery channel $\Lambda_b \to J/\psi \, K^- \! p$,
  namely, $\Lambda_c \to \phi \pi^0 \! p$, is feasible for this study
  and indeed is less Cabibbo-suppressed, although its limited phase
  space suggests that evidence of a $s\bar s uud$ resonance $P_s^+$
  would be confined to the kinematic endpoint region.
\end{abstract}


\maketitle


\section{Introduction} \label{sec:Intro}

The discovery of multiple exotic charmoniumlike (and bottomoniumlike)
states in the past dozen years, starting with Belle's
observation~\cite{Choi:2003ue} of the $X(3872)$, has been nothing
short of stunning.  Previously, the heavy quarkonium $Q\bar Q$ systems
were the best understood sectors of hadronic physics, with spectra
that could be completely predicted from two-body Schr\"{o}dinger
equations, modeled using nonrelativistic potentials developed decades
ago and continuously
refined~\cite{Eichten:1978tg,Eichten:1979ms,Barnes:2005pb}.  The
observation of more than 20 such so-called $X,Y,Z$ bosonic
states~\cite{Brambilla:2014jmp} (several of which have been confirmed
by multiple experiments) has upended this previous simple picture.
The latest finding in this regard is the LHCb
observation~\cite{Aaij:2015tga} of two exotic baryonic charmoniumlike
states, $P_c^+(4380)$ and $P_c^+(4450)$, at high statistical
significance in the $J/\psi \, p$ spectrum of $\Lambda_b \to J/\psi \,
K^- p$.  These states, like the $Z(4475)$ [also called $Z(4430)$]
before them~\cite{Aaij:2014jqa}, are shown to have rapid phase
variation in their production amplitudes consistent with true resonant
behavior.  Evidence continues to mount that at least some of $X$, $Y$,
and $Z$ are genuine $c\bar c q_1 \! \bar q_2$ tetraquark states (not
just kinematical effects), while the $P_c^+$ are $c\bar c uud$
pentaquark states.

Of course, the possibility of QCD exotics has been noted in the
earliest days of the quark model~\cite{GellMann:1964nj,Zweig:1964jf},
even before the advent of color dynamics.  Nevertheless, despite
decades of scrutiny, no unambiguous experimental signal indicating the
existence of an exotic hadron outside of the $q\bar q$-meson,
$qqq$-baryon paradigm has ever been identified in the light-quark
$(u,d,s)$ sector, or even in the sector with a single heavy quark.
Why should it be that exotics are first becoming visible in doubly
heavy-quark systems?  One explanation lies in the embarrassment of
hadronic riches in the $< 2.5$~GeV range: Many of the purported
light-quark exotics have the same quantum numbers as conventional
quark-model states and can hide amongst them, or indeed, mix with them
quantum mechanically.  Even the extremely well-established $J^{PC} =
1^{++}$ $X(3872)$ might mix at some level with the yet-unseen
conventional $c\bar c$ state $\chi_{c1} (2P)$ (as suggested numerous
times, most recently in Ref.~\cite{Achasov:2015oia}).

Previous work by the present
author~\cite{Brodsky:2014xia,Blitz:2015nra,Brodsky:2015wza,
Lebed:2015tna,Lebed:2015fpa} to explain this curious fact has argued
that the key feature in forming an identifiable exotic state is the
presence of two components in the hadron, each of which contains a
heavy quark and is consequently fairly compact (a few tenths of a fm),
but that are separated from each other, in the sense of having a small
wave function overlap, by a somewhat larger distance.  The specific
proposal in those works is the presence of compact colored diquark (or
triquark) components separately bound together through the attractive
${\bf 3} \otimes {\bf 3} \supset \bar {\bf 3}$ color interaction and
collectively bound together by confinement.  However, the same
situation arises in the molecular picture, in which the compact
components are color-singlet heavy quark-containing meson and baryon
pairs.  In the former case, the residual color interaction between the
components is full-strength QCD, and in the latter it is the much
weaker residual color van der Waals QCD force.

Supposing that exotics have only recently become observable because
they require well-separated components, each containing a heavy quark,
one may reconsider analogous systems in which the $c\bar c$ or $b\bar
b$ quark pairs are replaced by $s\bar s$.  This proposal is not at all
guaranteed success, since the $s$ is not truly heavy ({\it i.e.}, the
current quark mass $m_s$ is smaller than $\Lambda_{\rm QCD}$), and it
may well turn out that exotics of the type thus far seen absolutely
require the presence of two heavy quarks.  A major thrust of this work
is to provide one test of this point of view: Is the divide between
the $c$ and $s$ quarks so great that {\em no\/} exotic behavior
survives in the $s\bar s$ sector?  In order to do so, one may seek out
effects perhaps not as prominent as in the heavy-quark systems since
$s$ is not truly heavy, but anomalous nonetheless.  This proposal was
first advocated and applied to the case of $\phi$-$N$ photoproduction
in Ref.~\cite{Lebed:2015fpa}, where it was used to explain the
appearance in CLAS (JLab) data~\cite{Dey:2014tfa,Dey:2014npa} of
peculiar enhancements of the $\gamma p \to \phi p$ cross section in
the $\phi$ forward and backward directions.  It was argued that
treating the process as a $2 \to 2$ scattering resulting from the
formation of a color-antitriplet $(su)$ diquark and a color-triplet
$[\bar s (ud)]$ antitriquark, which subsequently hadronize to $\phi \
[ = \! (\bar s s)]$ and $p \ [= \!  u(ud)]$ through the
large-separation wave function tails of the hadrons stretching between
the two compact colored components, provides a natural mechanism to
explain forward and backward cross section enhancements.  In this
case, the anisotropic nature of the enhancements disfavors a resonant
origin~\cite{Dey:2014npa}, meaning that the $(su)$-$[\bar s (ud)]$
complex is only a ``would-be'' pentaquark~\cite{Lebed:2015fpa}.

The original $\phi$ photoproduction study was suggested by the simple
substitution of $c\bar c \to s\bar s$ into the $\gamma N \to P_c \to
J/\psi \, N^{(*)}$ photoproduction proposals of
Refs.~\cite{Wang:2015jsa,Kubarovsky:2015aaa,Karliner:2015voa}.  In
both $\phi$ and $J/\psi$ cases, the $Q\bar Q$ pair arises through the
dissociation of the incoming photon.  No particular $P_c$
compositeness substructure
(diquark~\cite{Maiani:2015vwa,Anisovich:2015cia,Ghosh:2015ksa,
  Anisovich:2015zqa,Wang:2015wsa} or other colored
substructure~\cite{Mironov:2015ica},
molecular~\cite{Chen:2015loa,Chen:2015moa,Roca:2015dva,He:2015cea,
  Meissner:2015mza,Huang:2015uda},\footnote{For predictions of
  hidden-charm baryons prior to the $P_c^+$ observation, see
  Refs.~\cite{Wu:2010jy,Wu:2010vk,Wang:2011rga,Yang:2011wz,Yuan:2012wz,
    Xiao:2013yca, Karliner:2015ina}.}
hadrocharmonium~\cite{Kubarovsky:2015aaa}, or
soliton~\cite{Scoccola:2015nia}) is presupposed; but if the states
turn out to be kinematical
effects~\cite{Guo:2015umn,Liu:2015fea,Mikhasenko:2015vca} due to the
particular placement of hadronic levels with respect to the original
$\Lambda_b \to J/\psi \, K^- \! p$ production process, then one would
not necessarily expect interesting structure to arise in $\gamma N \to
J/\psi \, N^{(*)}$.  A discussion of the relative merits of various
interpretations for $P_c^+$ states appears in
Ref.~\cite{Burns:2015dwa}.

Investigations have also begun into decays related to $\Lambda_b \to
J/\psi \, K^- \! p$, such as via $\Xi_b$ and
$\Omega_b$~\cite{Li:2015gta,Cheng:2015cca,Chen:2015sxa}, in a search
for the flavor SU(3) partners of the $P_c^+$ states.  In each case,
the underlying weak decay is $b \to c W^{*-} \to c \bar c s$ or $b \to
c W^{*-} \to c \bar c d$, meaning that the relevant
Cabibbo-Kobayashi-Maskawa (CKM) matrix element combination is
$V^{\vphantom{*}}_{cb} V^*_{cs\vphantom{b}}$ or the even smaller
combination $V^{\vphantom{*}}_{cb} V^*_{cd}$.  The possibility that
the weak decay is actually $b \to u \bar u s$ is explored in
Ref.~\cite{Hsiao:2015nna}.

\begin{figure}[t]
\begin{center}
\includegraphics[width=\linewidth]{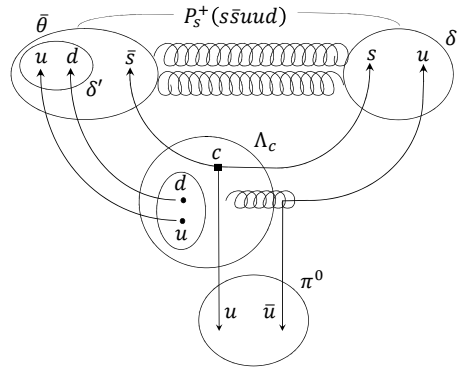}
\caption{Illustration of one particular mechanism [diquark
  ($\delta$)-antitriquark ($\bar \theta$) formation] for the
  production of a pentaquark $P_s^+ = s\bar s uud$ state in the decay
  $\Lambda_c \to P_s^+ \pi^0$.  The black square indicates the
  $c$-quark weak decay.  The spatially extended $\delta \bar \theta$
  state is held together by long-range color forces (indicated by
  gluon lines) via a color flux tube.}
\label{Fig:Ps}
\end{center}
\end{figure}

In this short paper we propose to create the precise
hidden-strangeness analogues $P_s^+ = s \bar s uud$ to the $P_c^+$
states through the decay $\Lambda_c \to P_s^+ \pi^0 \to \phi \, \pi^0
p$.  The replacement of $b \to c$, specifically the substitution of
the weak decay $c \to s W^{*+} \to s \bar s u$, is all that is needed
to produce this channel from one already known.  We illustrate in
Fig.~\ref{Fig:Ps} the flow of quark flavors in the process in terms of
the diquark-triquark picture used for $P_c^+$ in
Ref.~\cite{Lebed:2015tna}; however, as argued above, the diquark
picture is not necessarily the only one that produces viable
double-heavy pentaquark states, and the figure is intended merely as
an illustration of one particular viable physical process.  Again, we
emphasize that $s\bar s$-containing exotics, and the $P_s^+$ states in
particular, are not guaranteed to exist; the proposal here is that the
underlying mechanism creating the $P_c^+$ states also holds for
$P_s^+$, and the sole dynamical input is the assumption that the
gluodynamics leading to the energy differences involved is flavor
independent.

A discussion of the merits and drawbacks of this and related modes is
presented in Sec.~\ref{sec:Modes}, followed by a brief summary in
Sec.~\ref{sec:Concl}.

\section{Decay Modes for Producing Hidden-Strangeness Pentaquarks}
\label{sec:Modes}

We begin with the analogy between the process $\Lambda_b \to P_c^+ K^-
\to J/\psi \, K^- p$ and $\Lambda_c \to P_s^+ \pi^0 \to \phi \, \pi^0
p$, the former illustrated in Ref.~\cite{Lebed:2015tna} and the latter
here in Fig.~\ref{Fig:Ps}.  Just as for the corresponding $P_c^+$
production process in Ref.~\cite{Lebed:2015tna}, the presence of more
than one copy of the light $u$ quark in the final state (indeed, three
$u$'s for $P_s^+ \pi^0$ compared to only two for $P_c^+ K^-$) offers
the possibility of several production diagrams, regardless of the
physical process.  In the diquark picture, it is natural to expect the
$u$ in the diquark $\delta^\prime$ contained in the initial
$\Lambda_{b,c}$ to maintain its identity throughout the process as a
spectator.  However, the $u$ appearing in the diquark $\delta$ can
still emerge either from the weak decay $c \to s \bar s u$ or from the
$u\bar u$ production.  Analogous statements apply to other physical
decay pictures.

Replacing $u\bar u \to d\bar d$ produces the isospin-partner process
$\Lambda_c \to P_s^0 (s\bar s udd) \, \pi^+ \to \phi \, \pi^+ \! n$.
The rate for this process should be comparable to that for $P_s^+$,
but not exactly equal (even in the limit of perfect isospin), since
the $c \to s \bar s u$ weak decay violates isospin, failing to produce
a $d$ quark able to interfere with the ones from other
sources.\footnote{The $d$ quark reemerges if one considers the weak
  decay $c \to d \bar s u$, but such processes are not only CKM
  suppressed (by a factor $|V_{cd}^*/V_{cs}^*|^2 \simeq
  \frac{1}{20}$), but also may lack an $s$ quark in the final state.}
In any case, neutron reconstruction can be experimentally rather
challenging, so the $P_s^+$ channel seems to be the most promising one
for near-term investigations.

To be precise, if the mixing associated with the weak-decay $u$ quark
is ignored, the processes $\Lambda_b \to P_c^+ K^- \to J/\psi \, K^-
p$ and $\Lambda_c \to P_s^+ \pi^0 \to \phi \, \pi^0 p$ are entirely
comparable if one substitutes $V^{\vphantom{*}}_{cb}
V^*_{cs\vphantom{b}} \to V^*_{cs} V^{\vphantom{*}}_{us}$ and notes
that $K^-$ and $\pi^0$ are SU(3)-partnered pseudo-Nambu--Goldstone
bosons.  Only the $\frac{1}{\sqrt{2}}$ in the $\pi^0$ flavor wave
function is different.

Noting that $|V^{\vphantom{*}}_{cb} V^*_{cs\vphantom{b}}|^2 /
|V^*_{cs} V^{\vphantom{*}}_{us}|^2 \simeq \frac{1}{30}$, one sees that
the corresponding charmed decays are substantially less CKM suppressed
than their bottom counterparts.  This simple fact explains why many of
the lighter $b$-hadrons have lifetimes comparable to their charmed
counterparts despite a much greater available phase space.
Conversely, the branching fractions for the corresponding charmed
processes such as $\Lambda_c \to P_s^+ \pi^0 \to \phi \, \pi^0 p$, or
for that matter, nonresonant $\Lambda_c \to \phi \, \pi^0 p$, are
enhanced compared to their $b$ counterparts.  A glance at the known
branching fractions for $\Lambda_c$ decays~\cite{Agashe:2014kda} shows
known three-body decays to occur in the several times $10^{-3}$ range
or even larger, indicating that, if nothing else, the discovery of the
mode $\Lambda_c \to \phi \, \pi^0 p$ should be straightforward.

The resonance $P_s^+$ content of the process is also expected to be
nonnegligible.  In the $P_c^+$ case, LHCb measured~\cite{Aaij:2015tga}
the branching fraction ratios
\begin{eqnarray}
\frac{{\rm B.R.} (\Lambda_b \to P_c^+ K^-)}
{{\rm B.R.} (\Lambda_b \to J/\psi K^- p)} = \left\{
\begin{array}{c}
(8.4 \pm 0.7 \pm 4.2) \% \\
(4.1 \pm 0.5 \pm 1.1) \%
\end{array}
\right. \, ,
\end{eqnarray}
for $P_c^+(4380)$ and $P_c^+(4450)$, respectively, quite significant
considering the large available $\Lambda_b$ phase space, suggesting
that $P_s^+$ production in $\Lambda_c \to \phi \, \pi^0 p$ will not be
uncommon.

Arguably, the most interesting difference between $\Lambda_b$ $\to
P_c^+ K^- \to J/\psi \, K^- p$ and $\Lambda_c \to P_s^+ \pi^0 \to \phi
\, \pi^0 p$ is simple phase space.  Since $m_b - m_c$ is so much
greater than $m_c - m_s$, a much greater phase space is available in
the former process.  To be specific,
\begin{eqnarray}
  m_{\Lambda_b} - m_{J/\psi} - m_p - m_{K^-} & = & 1090.64 \pm 0.23
  \ \rm{MeV} \, , \nonumber \\
  m_{\Lambda_c} - m_\phi - m_p - m_{\pi^0} & = & \ \, 193.75 \pm 0.14
  \ \rm{MeV} \, . \nonumber \\ & & \label{eq:thresh1}
\end{eqnarray}
Note first that the phase space for the $\Lambda_c$ decay is so small
that no unflavored meson in the final state heavier than $\pi$, and no
unflavored baryon heavier than a nucleon, is possible.  In the case of
final-state decays to states in which the heavier quarks emerge in
separate hadrons:
\begin{eqnarray}
  m_{\Lambda_b} - m_{\overline{D}^{*0}} - m_{\Lambda_c} - m_{K^-} & =
  & 832.40 \pm 0.28 \ \rm{MeV} \, , \nonumber \\
  m_{\Lambda_c} - m_{K^{*+}} - m_\Lambda - m_{\pi^0} & = & 144.14
  \pm 0.29 \ \rm{MeV} \, , \nonumber \\ & & \label{eq:thresh2}
\end{eqnarray}
which assumes that the mesons appearing in the $P_{c,s}^+$ resonance
decays preferentially have $J^P = 1^-$ like $J/\psi$ or $\phi$, then
the phase space is even smaller.  Obviously, if $\bar D^0$ or $K^+$
final states can occur in the $P_c^+$ or $P_s^+$ decays, respectively
(which requires higher partial waves since the $P_c^+$ states have $J
= \frac 3 2$, $\frac 5 2$), then the phase space is correspondingly
larger.  But the message is clear: If $P_s^+$ resonances are formed in
the decay of $\Lambda_c$, they do not have much available phase space.

This effect is further magnified if one notes that the observed
$P_c^+$ resonances lie well above the $J/\psi \, p$ threshold, by
about 415 and 345~MeV for $P_c(4450)$ and $P_c(4380)$, respectively.
If similar numbers hold for the distance of the purported $P_s^+$
states from the $\phi \, p$ threshold, which assumes a flavor
independence of the mechanism depicted in Fig.~\ref{Fig:Ps} (an
extremely crude first approximation), then according to
Eqs.~(\ref{eq:thresh1})--(\ref{eq:thresh2}), the resonance peaks will
not be visible in $\Lambda_c$ decays.  Indeed, the $P_c(4450)$ is
sufficiently narrow ($\Gamma = 39 \pm 20$~MeV) that a corresponding
$P_s^+$ resonance (peak at 2372~MeV) would not be visible at all.
However, the large width of $P_c(4380)$ ($\Gamma = 205 \pm 88$~MeV)
suggests that an exact $P_s^+$ analogue (peak at 2303~MeV) would begin
to appear in the endpoint region.  The topology of the event in the
center-of-momentum (c.m.)  frame of the $\Lambda_c$ in such a case
would be remarkable: Since this c.m.\ frame is the rest frame of not
only the $\Lambda_c$ but almost that of the $P_s^+$ and $\pi^0$ as
well, one would observe two photons of nearly equal energy from the
$\pi^0$ decay emerge back-to-back, as well as a slowly moving proton
recoiling against a nearly collinear $K\bar K$ pair from the $\phi$
decay.

The higher-strangeness SU(3) partners to the decay $\Lambda_c \to \phi
\, \pi^0 p$ whose parent baryon decays weakly, namely $\Xi_c^+ \to
\phi \, \overline{K}^0 p$ and $\Omega_c \to \phi \, \overline{K}^0
\Lambda$, tend to have even less phase space: about 13 and 62~MeV,
respectively.  However, the decay $\Xi_c^+ \to \phi \, \pi^+ \Lambda$
has almost precisely the same phase space as $\Lambda_c \to \phi \,
\pi^0 p$:
\begin{equation}
m_{\Xi_c^+} - m_\phi - m_\Lambda - m_{\pi^+} = 193.22 \pm 0.40 \ {\rm
  MeV} \, .
\end{equation}
Not only does the $\Xi_c^+$ have a longer lifetime than $\Lambda_c$
($\tau_{\Xi_c^+} = 0.44$~ps, to be compared with $\tau_{\Lambda_c} =
0.20$~ps or $\tau_{\Lambda_b} = 1.47$~ps), it has a perhaps more
easily reconstructed final state: a primary decay $\pi^+$ and a
secondary decay $\Lambda \to p \pi^0$.  Note that the channels with
$\phi \Lambda$ in the final state imply open-strangeness ($s \bar s
uds$) pentaquarks, which may be expected to lie above $P_s^+ = s \bar
s uud$ states in mass, and therefore potentially outside of range of
the initial hadron phase space.  It is also worth noting that the
$\Xi_c^+ = cus$ and $\Omega_c = css$ contain diquarks of a somewhat
different nature than the $(ud)$ in a $\Lambda$ state, the former
carrying nonzero isospin and the latter requiring proper
antisymmetrization between the identical $s$ quarks.  Indeed, part of
the motivation for this work (see Fig.~\ref{Fig:Ps}) was to treat the
$s$ quarks as heavy, which becomes less compelling if $s$ quarks arise
elsewhere in the process.

The small available phase space remains the least appealing feature of
this proposal.  Even so, several points are worth mentioning.  First,
the $P_s^+$ mass estimates may be unnecessarily high.  Since the
observed $P_c^+$ states have the large spins $J = \frac 3 2$ and
$\frac 5 2$ (as well as one of them necessarily having negative
parity), it is very likely that lighter yet-unobserved $P_c^+$ states
exist, which implies lighter $P_s^+$ would be possible as well.
Indeed, no obvious physical principle requires the $P_s^+$ states to
lie above the $\phi \, p$ threshold as far as the corresponding
$P_c^+$ states lie above the $J/\psi \, p$ threshold.  Second, the
large available phase space for $P_c^+$ decay was actually a
substantial nuisance in the $P_c^+$ observation
paper~\cite{Aaij:2015tga}, since multiple excited $\Lambda$ states had
to be included in the analysis.  With phase space as small as
suggested above, potential higher states provide very little
contamination.

One may hope to avoid the phase-space problem, as well as reduce the
CKM suppression, by considering not $c \to s W^{*+} \to s \bar s u$
but $c \to s W^{*+} \to s \bar d u$, which increases the production
rate of relevant processes by a factor of $|V_{ud}/V_{us}|^2 \simeq
20$.  However, the $P_s^+$ state still requires an $\bar s$ quark,
which must now appear through pair production.  The final state then
requires one to accommodate at least three strange quarks, creating
insurmountable difficulties with phase space in charmed baryon decays.
Alternately, one may attempt to build states in the manner of
Fig.~\ref{Fig:Ps} without an $\bar s$ quark in the antitriquark $\bar
\theta$, but doing so violates the premise of exotics requiring a
heavy quark in each component to be detectable.

\section{Conclusions} \label{sec:Concl}

A search for the hidden-strangeness pentaquarks $P_s^+ = s\bar s uud$,
siblings to the newly observed $P_c^+$ pentaquark candidates, appears
to be well within current experimental capabilities.  The main
difficulty appears to be the limited phase space available in the
decays of the likely charmed baryon sources, the best candidate being
$\Lambda_c \to P_s^+ \pi^0 \to \phi \, \pi^0 p$, which is the exact
SU(3)-flavor analogue to the channel $P_c^+$ discovery channel
$\Lambda_b \to J/\psi \, K^- \! p$.  The decay $\Xi_c^+ \to \phi
\Lambda \pi^+$ appears to be interesting both for the relative ease of
its reconstruction and the possibility of finding hints of an
open-strangeness $s \bar s uds$ pentaquark.  At minimum, new
unobserved decay modes are within reach, but with a bit of luck, more
exciting hints of exotic hadron structure may be uncovered.

\begin{acknowledgments}
  I thank P.~Koppenburg for useful discussions about the remarkable
  capabilities of LHCb. This work was supported by the National
  Science Foundation under Grant Nos.\ PHY-1068286 and PHY-1403891.
\end{acknowledgments}


\end{document}